\documentclass[sigconf]{acmart}

\usepackage{svg}
\usepackage{subcaption}

\AtBeginDocument{%
  \providecommand\BibTeX{{%
    \normalfont B\kern-0.5em{\scshape i\kern-0.25em b}\kern-0.8em\TeX}}}

\copyrightyear{2024}
\acmYear{2024}
\setcopyright{acmlicensed}
\acmConference[IH\&MMSec '24]{Proceedings of the 2024 ACM Workshop on Information Hiding and Multimedia Security}{June 24--26, 2024}{Baiona, Spain}
\acmBooktitle{Proceedings of the 2024 ACM Workshop on Information Hiding and Multimedia Security (IH\&MMSec '24), June 24--26, 2024, Baiona, Spain}
\acmDOI{10.1145/3658664.3659656}
\acmISBN{979-8-4007-0637-0/24/06}

\begin{document}

\title{Is Audio Spoof Detection Robust to Laundering Attacks?}


\author{Hashim Ali}
\email{alhashim@umich.edu}
\orcid{0000-0003-0532-0268}
\affiliation{%
  \institution{University of Michigan-Dearborn}
  \streetaddress{4901 Evergreen Rd}
  \city{Dearborn}
  \state{MI}
  \country{USA}
  \postcode{48128}
}

\author{Surya Subramani}
\authornote{Authors contributed equally to this research.}
\email{suryasss@umich.edu}
\orcid{0009-0007-2394-8901}
\affiliation{%
  \institution{University of Michigan-Dearborn}
  \streetaddress{4901 Evergreen Rd}
  \city{Dearborn}
  \state{MI}
  \country{USA}
  \postcode{48128}
}

\author{Shefali Sudhir}
\authornotemark[1]
\email{sgokarn@umich.edu}
\orcid{0009-0003-4609-0734}
\affiliation{%
  \institution{University of Michigan-Dearborn}
  \streetaddress{4901 Evergreen Rd}
  \city{Dearborn}
  \state{MI}
  \country{USA}
  \postcode{48128}
}

\author{Raksha Varahamurthy}
\authornotemark[1]
\email{rakshav@umich.edu}
\orcid{0009-0002-6310-8767}
\affiliation{%
  \institution{University of Michigan-Dearborn}
  \streetaddress{4901 Evergreen Rd}
  \city{Dearborn}
  \state{MI}
  \country{USA}
  \postcode{48128}
}

\author{Hafiz Malik}
\email{hafiz@umich.edu}
\orcid{0000-0001-6006-3888}
\affiliation{%
  \institution{University of Michigan-Dearborn}
  \streetaddress{4901 Evergreen Rd}
  \city{Dearborn}
  \state{MI}
  \country{USA}
  \postcode{48128}
}


\renewcommand{\shortauthors}{Hashim Ali, Surya Subramani, Shefali Sudhir, Raksha Varahamurthy \& Hafiz Malik}

\begin{abstract}

Voice-cloning (VC) systems have seen an exceptional increase in the realism of synthesized speech in recent years. The high quality of synthesized speech and the availability of low-cost VC services have given rise to many potential abuses of this technology. Several detection methodologies have been proposed over the years that can detect voice spoofs with reasonably good accuracy. However, these methodologies are mostly evaluated on clean audio databases, such as ASVSpoof 2019. This paper evaluates SOTA Audio Spoof Detection approaches in the presence of laundering attacks. In that regard, a new laundering attack database, called ASVSpoof Laundering Database, is created. This database is based on the ASVSpoof 2019 (LA) eval database comprising a total of 1388.22 hours of audio recordings. Seven SOTA audio spoof detection approaches are evaluated on this laundered database. The results indicate that SOTA systems perform poorly in the presence of aggressive laundering attacks, especially reverberation and additive noise attacks. This suggests the need for robust audio spoof detection.

\end{abstract}



\maketitle

\section{Introduction} \label{intro} 
The last few years have seen an exceptional increase in the realism of synthesized speech. Conceivably, the most prominent development is zero-shot, multi-speaker text-to-speech (ZS-TTS) \cite{casanova2022yourtts, wang2301neural, li2023zse} using which it is possible to synthesize voice for speakers not seen during training, with only a few seconds to minutes of reference audio. This advancement in TTS technology enabled the formation of numerous commercial entities providing low-cost TTS services to their users, such as ElevenLabs.

Due to the high quality of synthesized speech, VC technology has promising applications in various areas of life. These applications range from cloning voices for people with speech impairments, cloning an actor's voice for dubbing or character portrayal, to cloning a voice for building digital avatars. Recently, a jailed politician in Pakistan, Imran khan, used deepfake technology to hold an online campaign rally featuring his AI-generated video addressing his supporters and urging them to vote in large numbers \cite{ray_imran_nodate}. While there are numerous benefits of VC technology, the potential of their abuse cannot be ignored. 

\subsection{Emerging Threats to audio spoof detection approaches} \label{emerg_threat}

In January 2024, around 25000 voters in New Hampshire received a deepfake robocall impersonating President Joe Biden, telling them to not vote in the state's primary elections. This robocall was analyzed by a security company, called Pindrop, and it was attributed to be likely generated through Elevenlabs' technology \cite{ knibbs_researchers_nodate}. The audio generated through Elevenlabs usually has very high quality; however, the Biden robocall was particularly noisy, which, if the audio actually came from Elevenlabs, hints towards the deliberate addition of noise in the audio to bypass audio spoof detection. 


The audio spoof detection methodologies described in Section \ref{BenchASD} have predominantly been evaluated on ASVSpoof (2015, 2017, 2019, 2021) datasets \cite{wu_asvspoof_2015, delgado_asvspoof_2018, wang_asvspoof_2020, yamagishi_asvspoof_2021}. With the exception of the ASVSpoof 2021 dataset, these corpora have been curated within controlled settings which may not accurately depict conditions encountered in real-world scenarios. For instance, ASVSpoof 2019 employs the VCTK corpus \cite{yamagishi_cstr_2019}, a multi-speaker English speech database that was compiled within the confines of a semi-anechoic chamber. 

In practical deployment scenarios, a speaker verification system can be subjected to complex auditory environments characterized by reverberations, mechanical noise, or conversations in the background. In the context of forensic analysis, an audio could have gone through multiple compression cycles, bear digital artifacts from social media platforms, or have been intentionally modified to circumvent spoof detection mechanisms—a notable illustration being the Biden robocall incident. Consequently, it is imperative for audio spoof detection methodologies to undergo rigorous evaluation using databases that encapsulate a spectrum of laundering attacks, closely mirroring real-world conditions.

To benchmark SOTA audio spoof detection approaches in real-world settings, this paper presents the following contributions: 

\begin{itemize}
    \item A new database based on ASVSpoof 2019 logical access (LA) eval partition \cite{wang_asvspoof_2020} is introduced. We call this database ``ASVSpoof Laundering Database''\textcolor{blue}{\footnote{https://issf.umd.umich.edu/downloads/data}}. The ASVSpoof 2019 LA eval database is passed through five different types of additive noise, three types of reverberation noise, six different re-compression rates, four different resampling factors, and one type of low pass filtering accumulating to a total of 1388.22 hours of audio data.
    
    \item We evaluate and benchmark seven SOTA audio spoof detection approaches on each type of laundering attack in ASVSpoof Laundering Database.

    \item We demonstrate that SOTA systems perform poorly in the presence of laundering attacks which suggest the need for more robust audio spoof detection for practical applications.
\end{itemize}

A brief review of the relevant literature in the space of robustness evaluation of audio spoof detection systems is provided in Section \ref{related}. After that, we discuss the bechnmarked audio spoof detection approaches from three categories in section \ref{BenchASD}. Next, laundering attacks to SOTA audio spoof detection approaches are discussed in section \ref{laundering}. Experimental setup and results are discussed in Sections \ref{exp_setup} and \ref{results}, respectively.

\section{Related Work} \label{related}
In 2016, two articles were published at the same time \cite{hanilci_spoofing_2016, tian_investigation_2016}. These articles studied the performance of audio spoof detection approaches of the time under additive and reverberation noise. Hanilci et al. \cite{hanilci_spoofing_2016} corrupted the original ASVSpoof 2015 signals with three additive noise types and evaluated eight feature sets with a common Gaussian Mixture Model (GMM) Classifier. The subband spectral centroid magnitude coefficient (SCMC) features tend to outperform other features for babble and car noises whereas Mel-frequency cepstral coefficient (MFCC) features perform better under white noise. Similarly, Tian et al. \cite{tian_investigation_2016} also generated a noisy database based on the ASVSpoof 2015 database. Five types of additive noise, and reverberation noise with three reverberant times RT60 $\in$ (0.3, 0.6, 0.9) s were considered in this work. The authors evaluated six types of feature sets on this noisy database with a common multilayer perceptron (MLP)-based classifier. In general, the magnitude-based features, log magnitude spectrum (LMS) and residual log magnitude spectrum (RLMS) perform worse than the phase-based features, instantaneous frequency derivative (IF), baseband phase difference (BPD), group delay (GD) and modified group delay (MGD). 

Muller et al. \cite{muller_does_2022} re-implemented twelve of the most popular architectures trained on ASVSpoof 2019 database and evaluated them on an in-the-wild audio deepfake dataset. The authors concluded that end-to-end models perform better than feature-based models, obtaining up to 1. 2\% EER on the ASVSpoof data set and 33. 9\% EER on the in-the-wild data set (Rawnet2 \cite{tak2021end} and RawGAT-ST \cite{jung2022aasist}). Spectrogram-based models perform slightly worse, achieving up to 6.3\% EER on ASVSpoof data and 37.4\% EER on in-the-wild dataset (LCNN \cite{lavrentyeva2019stc}).

Several studies have explored the efficacy of audio spoof detection systems for speech compression. The ASVSpoof 2021 Logical Access (LA) task aimed at gauging the resilience of these systems to nuisance variations caused by compression, packet loss, and various other distortions, resulting in the creation of seven distinct testing scenarios \cite{yamagishi_asvspoof_2021}. Similarly, the ASVSpoof 2021 deepfake (DF) task introduced several lossy compression codecs typically used to store media, generating a total of 9 different evaluation conditions. Moreover, Zhang et al. \cite{zhang2023compressed} and Yadav et al. \cite{yadav2023assd} conducted research to assess the performance of audio spoofing detection on compressed speech formats prevalent in social networking environments, including MP3 and Advanced Audio Coding (AAC).

The performance of audio spoof detection in acoustically degraded conditions (laundering attacks) has not received adequate attention. Hanilci et al. \cite{hanilci_spoofing_2016} and Tian et al. \cite{tian_investigation_2016} conducted their experiments on the ASVSpoof 2015 data set. Thereafter, significant advancements in VC and TTS technology have made it possible to produce high-quality synthesised audio (section \ref{intro}). Subsequently, improved audio spoof detection techniques were introduced, particularly end-to-end learning systems that exhibit significantly superior performance compared to traditional feature-based systems. Furthermore, ASVSpoof 2021 \cite{yamagishi_asvspoof_2021}, Zhang et al. \cite{zhang2023compressed}, Yadav et al. \cite{yadav2023assd} and the proposed work differ in that the latter also focuses on pre-sensor measurement attacks, such as additive noise, reverberation, and low-pass filtering attacks, while the former concentrates on post-sensor attacks, i.e., transmission or compression artifacts introduced in the audio signal after the audio is captured. Although Muller et al. \cite{muller_does_2022} performed a remarkable job in evaluating the performance of audio spoof detection on an in-the-wild dataset, we argue that the audio spoofs available online have undergone a number of post-processing steps, such as reverberation, recompression, and additive noise. As a result, an in-the-wild audio sourced from the internet could be a clean audio file that has been subjected to laundering attacks. As such, formalising SOTA systems' behaviour in the face of different types of laundering attacks is crucial. 

\section{Benchmarked audio spoof detection approaches} \label{BenchASD}

A significant amount of research has been done to develop strategies that can detect audio spoofs reliably. These strategies can be broadly classified into three categories \cite{ali2023protecting, khan2023battling, balamurali2019toward, kamble2020advances}, 
\begin{enumerate}
    \item Conventional Machine Learning Approaches
    \item Representation Learning Approaches
    \item End-to-end Learning Approaches 
\end{enumerate}

This section presents an overview of audio spoof detection approaches selected for this study. At least two audio spoof detection approaches from each category are selected. This selection is based on the availability of the open-source code. The description and implementation detail of each audio spoof detection approach grouped by category is given below, 


\subsection{Conventional Machine Learning Approaches} \label{ML}

Conventional machine learning (ML)-based approaches for audio spoof detection typically consist of two parts. The first part deals with hand-crafted feature extraction (front-end) and the second part consists of a model (back-end) that determines the authenticity of the audio signal \cite{balamurali2019toward, khan2023battling}.

Two audio spoof detection approaches, CQCC-GMM and LFCC-GMM are selected. Both approaches employ Gaussian Mixture Models (GMM) as back-end, and constant-Q cepstral coefficients (CQCC) and linear frequency cepstral coefficients (LFCC) features as front-ends. The configuration for both approaches is described in \cite{yamagishi_asvspoof_2021, liu_asvspoof_2023} and is kept the same as default. CQCC-GMM and LFCC-GMM report an EER of 8. 9\% and 3. 7\%, respectively, on the ASVSpoof 2019 LA eval database \cite{todisco19_interspeech}.

\subsection{Representation Learning Approaches} \label{representation}

Representation learning approaches work either in the form of feature learning \cite{qian2016deep} or as a pattern classifier \cite{yu2017spoofing}. Two approaches are also selected from this category. 

The first selected approach is LFCC-LCNN \cite{lavrentyeva2019stc}. In this approach Lavrentyeva et al. \cite{lavrentyeva2019stc} explored several types of acoustic features with a light convolutional neural network (LCNN) architecture. The authors reported that the lowest min-tDCF was achieved by LFCC-LCNN system with an EER of 5.06\% on the ASVSpoof 2019 LA dataset. The configuration details are provided in \cite{yamagishi_asvspoof_2021, liu_asvspoof_2023}. The second system is called OC-Softmax \cite{zhang2021one}. You and Jiang et al. formulated the voice anti-spoofing problem as a one-class classification problem. The key idea is to capture the target class distribution and set a tight boundary around it, so that all samples that belong to non-target class would fall outside this boundary. This method achieved an EER of 2.19\% on the ASVSpoof 2019 LA eval dataset.

\subsection{End-to-end Learning Approaches} \label{end-to-end}

End-to-end learning approaches for audio spoof detection operate directly upon raw waveform input, streamlining the training and evaluation process. Three audio spoof detection approaches are selected from this category.

The first system uses a modified RawNet2 architecture \cite{jung2020improved, tak2021end}, and reported inferior results compared to the baseline method, i.e. LFCC-GMM. The pooled EER for the base line was 3. 5\%, while the proposed RawNet2 architecture reported 5.13\%. The details of the implementation of this system are provided in \cite{yamagishi_asvspoof_2021, tak2021end, liu_asvspoof_2023}. The second system in this category is a direct extension to RawNet2, called RawGat. Tak et al \cite{tak21_asvspoof} proposed the use of graph attention networks (GATs) for the detection of audio spoofing attacks. RawGat-ST system outperformed all other baseline systems by a substantial margin and reported an EER of 1.06\% in the ASVSpoof 2019 LA database. The third system selected in this category is an extension to RawGat-ST, named AASIST \cite{jung2022aasist}. AASIST outperforms RawGAT-ST baseline with an EER of 0.083\% on the ASVSpoof 2019 LA eval database.

The implementation of all benchmarked systems is available at the following GitHub repository\textcolor{blue}{\footnote{https://github.com/hashim19/Rob-ASD}}


\begin{figure*}[!h]
\centering
\includegraphics[width=0.9\linewidth]{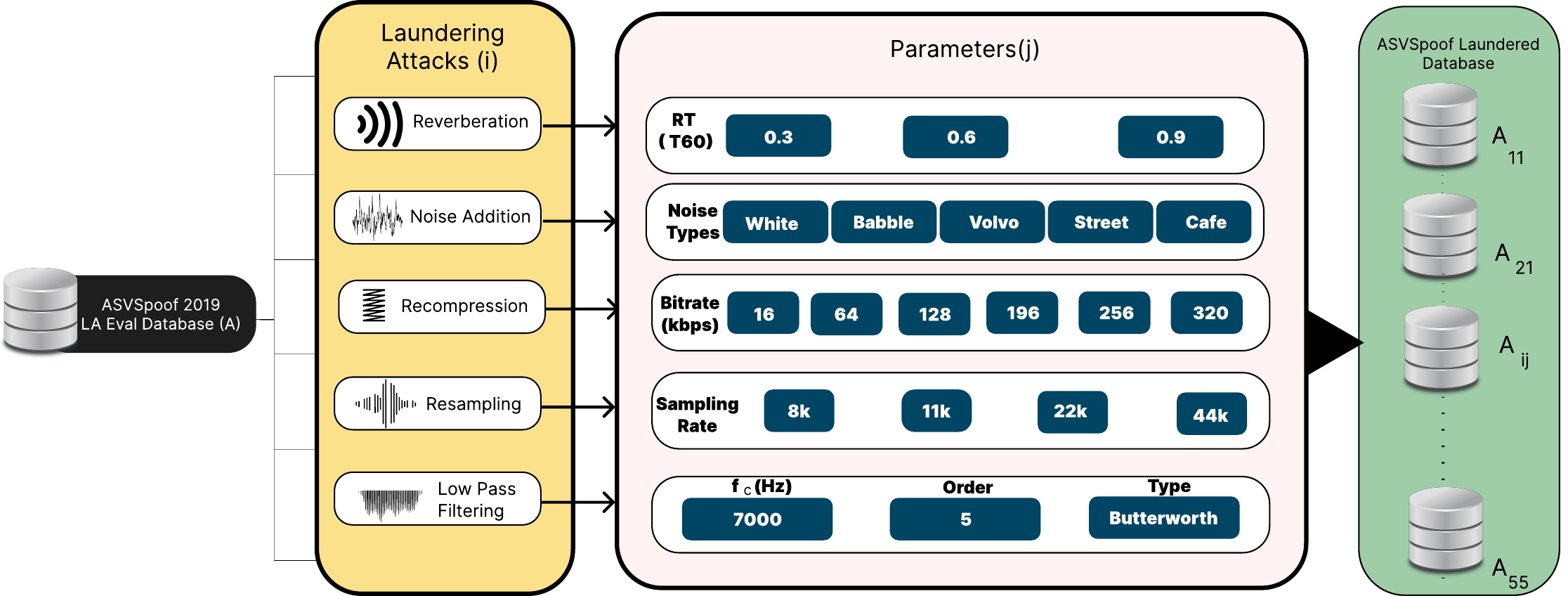}
\caption{Data Block Diagram: \normalfont ASVSpoof 2019 LA eval is the input database $A$. First column describes the laundering attacks (i). Second column describes the parameters (j) for each laundering attack. Third column describes the generated Laundered Database $A_{ij}$}
\label{fig:data_diagram}
\end{figure*}

\section{Laundering Attacks} \label{laundering}
In order to represent real-world settings for selected audio spoof detection systems, a new database based on the ASVSpoof 2019 LA eval database \cite{wang_asvspoof_2020} is proposed. This database is called ASVSpoof Laundered Database and is generated by adding various laundering attacks to the ASVSpoof 2019 LA eval database. Each type of laundering attack (i) with different parameters (j) is described in the figure \ref{fig:data_diagram} and in the following subsections.

\vspace{-1em}
\subsection{Reverberation}
Room reverberation refers to the persistence of sound in an enclosed space after the original sound source has stopped. It occurs because of reflections of sound waves from surfaces such as walls, ceilings, and floors. To add reverberation to an audio signal, we have used a library called pyroomacoustics \cite{scheibler2018pyroomacoustics}. First, a shoe-box room is simulated with a dimension of (10m, 7.5m, 3.5m), with source location and mic position of (2.5m, 3.7m, 1.76m) and (6.3m, 4.9m, 1.2m) respectively. After that, the reverberation time (RT60), i.e. the time it takes for Room Impulse Response (RIR) to decay by 60dB, is varied from 0.3s, 0.6s to 0.9s. This creates a total of 3 copies of the ASVSpoof 2019 LA eval database, one for each RT60.

\vspace{-1em}
\subsection{Additive Noise}
The noise addition process involves the introduction of controlled disturbances to data to simulate real-world conditions. This study involves five additive noises from two databases: White noise, Babble noise, Volvo noise from the SPIB database \textcolor{blue}{\footnote{https://web.archive.org/web/20120905015656/http://spib.rice.edu/spib/select\_noise.html}} and cafe noise, street noise from the QUT-NOISE database \cite{dean2015qut}. The detail of each type of noise as also described in\cite{tian_investigation_2016} is given below,

\begin{enumerate}

\item \textbf{White Noise:} A random noise with a consistent power spectral density.

\item \textbf{Babble Noise:} The sound of numerous conversations in a cafeteria setting with approx. 100 individuals speaking.

\item \textbf{Volvo Noise:} The internal noise within a Volvo 340 recorded while driving on a wet asphalt road.

\item \textbf{Street Noise:} Diverse ambient noise captured in an urban area, comprising traffic, pedestrian, and bird noises.

\item \textbf{Cafe Noise:} Assorted ambient noise recorded in a cafe environment, dominated by conversations and kitchen clatter.

\end{enumerate}

To add noise to clean audio signals, the Addshort Noise API from the Python library audiomentation \textcolor{blue}{\footnote{https://github.com/iver56/audiomentations?tab=readme-ov-file}} is used, which allowed us to add noise at SNR levels of 0dB, 10dB, and 20dB. This creates a total of 18 copies of the ASVSpoof 2019 LA eval database, one for each additive noise and SNR level.

\vspace{-1em}
\subsection{Recompression, Resampling and Low-Pass Filtering}
The audio files in ASVSpoof 2019 database are in FLAC (Free Lossless Audio Codec) format, with a bit-rate of 256 kbit/s. These audio files are first uncompressed to the WAV format. After that, the WAV files are compressed to MP3 format using bit rates of 16, 64, 128, 192, 256, and 320 kbit/s. Subsequently, all mp3 files are uncompressed to WAV format before passing to the selected systems. This process creates a total of six copies of the ASVSpoof 2019 LA eval database, one for each bit rate. We utilized ffmpeg for the recompression task.

The audio files in ASVSpoof 2019 database have 16 kHz sampling rate. These files are resampled with sampling rates of 8000 Hz, 11025 Hz, 22050 Hz, and 44100 Hz. This creates a total of four copies of the ASVSpoof 2019 LA eval database, one for each sampling rate. Moreover, the audio files from ASVSpoof 2019 LA eval database are passed through a low-pass butter-worth filter with a cut-off frequency of 8KHz and order 5 to generate exactly one low-pass filtered copy of the database. 

\section{Experimental Setup} \label{exp_setup}

\begin{table}
\centering
\caption{Statistics of the ASVSpoof 2019 Database}
\vspace*{-\baselineskip}
    \begin{tabular}{ | p{2cm} | p{1.2cm} p{1.2cm} | p{2cm} |}
     \hline
     Subset  & \multicolumn{2}{|c|}{Number of Utterances} &  Attacks \\
     & {Bonafide} & {Spoof} & \\
     \hline
     Train   &  2,580  &  22,800 & A01 - A06\\
     Development  &  2,548  &  22,296 & - \\
     Evaluation  &  7,355  &  63,882 & A06 - A19\\
     \hline
    \end{tabular}
\label{table:Asvspoof_2019} 
\vspace{-1em}
\end{table}

\begin{table}

\caption{Statistics of the ASVSpoof Laundered Database.\\ {\normalfont \textbf{Rev:} \normalfont Reverberation, \textbf{AN:} \normalfont Additive Noise, \textbf{Rec:} \normalfont Recompression, \textbf{Res:} \normalfont Resampling, \textbf{LPF:} \normalfont Low-pass filter}}
\vspace*{-\baselineskip}

    \begin{tabular}{| c | c c c c c|}
     \hline
     Partition Type  & \multicolumn{5}{|c|}{Laundering Type (Number of Utterances)} \\
     & {Rev} & {AN} & {Rec} & {Res} & {LPF} \\
     \hline
     Bonafide   &  22,065  &  132,390 & 44,130 & 29,420 & 7,355\\
     Spoof  &  191,646  &  1,149,876 & 383,292 & 255,288 & 63,88\\
     \hline
    \end{tabular}
    \label{table:Asvspoof_Launder} 
\vspace{-1em}
\end{table}

The details of the ASVSpoof 2019 database are given in table \ref{table:Asvspoof_2019}. It consists of three types of subsets, namely training, development, and evaluation. The training subset is used to train the audio spoof detection systems, and the evaluation subset is used to evaluate the trained audio spoof detection systems. The detail of the ASVSpoof Laundered Database is given in table \ref{table:Asvspoof_Launder}, and shows the number of bonafide and spoof utterances for each type of laundering attack.

The selected audio spoof detection systems are trained on clean ASVSpoof 2019 LA train database. This aligns with most relevant research, as well as the ASVSpoof 2019 Challenge evaluation process. After that these systems are evaluated on two databases, (i) ASVSpoof 2019 LA eval, (ii) ASVSpoof Laundered Database (refer to section \ref{laundering}, table \ref{table:Asvspoof_Launder}).

Following the ASVSpoof challenge evaluation plans, the equal error rate (EER) is used as an objective evaluation metric in our experiments \cite{wu_asvspoof_2015}. We omit the tandem detection cost function (min t-DCF) as it requires the false alarm and miss costs of the ASV system, which are only available for ASVSpoof 2019 eval database. EER corresponds to a CM operating point at which miss rate ($P_{miss}$) and False alarm rate ($P_{fa}$) becomes equal.

\section{Results} \label{results}

The results of our experiments are displayed in table \ref{table:Laundering_Res}, where selected systems are evaluated against all laundering attacks. The results for the ASVSpoof 2019 LA eval database are shown in table \ref{table:Laundering_Res} (row ASVSpoof19 Eval). We can observe that Conventional ML based systems perform worst with an EER of 8.9\% and 3.7\% for CQCC-GMM and LFCC-GMM respectively. Representation Learning systems perform better than conventional systems with an EER of 6.35\% and 5.8\% for LFCC-LCNN and OC-Softmax, respectively. The best performing systems are end-to-end learning systems with AASIST reporting the lowest EER of 0.83\%, followed by RawGat-ST and RawNet2 with an EER of 1.06\% and 4.6\%, respectively. In general, audio spoof detection systems exhibit similar performance on the ASVSpoof Laundered Database, that is, end-to-end learning systems perform better than representation learning systems, which perform better than conventional ML systems.

\begin{table*}
\centering
\caption{Results of the SOTA Audio Spoof Detection Approaches on Laundered Database shown in terms of EER (\%)}
\vspace*{-\baselineskip}
    \begin{tabular}{| c c | c c | c c | c c c |}
     \hline
     \multicolumn{2}{|c|}{}  & \multicolumn{2}{|c|}{Conventional} & \multicolumn{2}{|c|}{Representation} & \multicolumn{3}{|c|}{End-to-End} \\
     & & {CQCC-GMM} & {LFCC-GMM} & {LFCC-LCNN} & {OC-Softmax} & {RawNet2} & {RawGat-ST} & {AASIST} \\
     \hline
     
     ASVSpoof19 Eval &  &  8.9  &  3.7 & 6.35 & 5.8 & 4.6 & 1.06 & 0.83\\
     \hline
     Laundering Attack(i) & Parameter(j) & & & & & & & \\
     \hline
                   & 0.3  &  41.13  &  22.4 & 16 & \textbf{9.01} & 32.81 & 27.21 & 36.4\\
     Reverberation & 0.6  &  48.47  &  22.9 & 20.68 & \textbf{15.06} & 41.45 & 43.02 & 55.27\\
                   & 0.9  &  51.55  &  25.96 & 26.67 & \textbf{22.04} & 43 & 48.69 & 58.9\\
                   & avg  &  47.05  &  23.75 & 21.12 & \textbf{15.37} & 39.09 & 39.64 & 50.19\\
     \hline

                    & 0   &  28.74  &  27.27 & 34.77 & 34.25 & 26.97 & \textbf{24.84} & 33.39\\
     Babble Noise   & 10  &  28.56  &  28.5 & 18.39 & 19.16 & \textbf{8.44} & 20.78 & 17.65\\
                    & 20  &  30.69  &  24.61 & 9.53 & 11.24 & 4.76 & \textbf{1.7} & 2.33\\
                    & avg  &  34.78  &  26.79 & 20.9 & 21.55 & \textbf{13.39} & 15.77 & 17.79\\
     \hline

                    & 0   &  37.92  &  11.02 & 8.83 & 29.25 & \textbf{8.73} & 19.71 & 10.48\\
     Volvo Noise    & 10  &  27.53  &  13.41 & 7.17 & 22.05 & 5.76 & 5.91 & \textbf{5.52}\\
                    & 20  &  15.65 &  6.02 & 6.63 & 15.05 & 4.77 & \textbf{1.51} & 1.76\\
                    & avg  &  27.03 &  10.15 & 7.54 & 22.12 & 6.42 & 9.04 & \textbf{5.92}\\
     \hline

                    & 0   &  42.24  &  22.74 & 28.88 & 17.65 & \textbf{15.12} & 33.85 & 41.31\\
     White Noise    & 10  &  43.78  &  24.14 & 30.06 & 14.37 & \textbf{7.88} & 20.16 & 12.66\\
                    & 20  &  37.58  &  30.14 & 21.36 & 12.87 & 5.43 & \textbf{1.2} & 3.95\\
                    & avg  &  41.2  &  25.67 & 26.77 & 14.96 & \textbf{9.48} & 18.4 & 19.31\\
     \hline

                    & 0   &  38.22  &  43.21 & 27.46 & 31.49 & 24.23 & \textbf{23.97} & 38.48\\
     Cafe Noise     & 10  &  36.69  &  40.46 & 19.43 & 28.82 & \textbf{8.93} & 19.16 & 14.74\\
                    & 20  &  33.32  &  35.29 & 12.5 & 22.35 & 5.09 & \textbf{2.06} & 2.81\\
                    & avg  &  36.08  &  39.65 & 19.8 & 27.55 & \textbf{12.75} & 15.06 & 18.68\\
     \hline

                    & 0   &  \textbf{29.68}  &  46.67 & 30.03 & 33.84 & 32.47 & 32.03 & 40.08\\
     Street Noise   & 10  &  40.45  &  46.29 & 17.54 & 24.86 & \textbf{12.1} & 17.63 & 21.26\\
                    & 20  &  39.49  &  38.79 & 9.66 & 18.14 & 5.09 & \textbf{2.35} & 2.9\\
                    & avg  &  36.54  &  43.92 & 19.08 & 25.61 & \textbf{16.55} & 17.34 & 21.41\\
     \hline

                    & 16  &  42.88  &  55.44 & 15.09 & 17.12 & 4.51 & 2.01 & \textbf{1.6}\\
                    & 64  &  13.68  &  33.8 & 6.26 & 15.29 & 5.09 & 1.09 & \textbf{0.88}\\
    Recompression   & 128  &  13.43  &  33.22 & 6.35 & 15.09 & 4.06 & 1.07 & \textbf{0.83}\\
                    & 192  &  13.4  &  33.25 & 6.35 & 15.1 & 4.07 & 1.07 & \textbf{0.83}\\
                    & 256  &  13.4  &  33.25 & 6.35 & 15.1 & 4.07 & 1.07 & \textbf{0.83}\\
                    & 320  &  13.4  &  33.25 & 6.35 & 15.1 & 4.07 & 1.07 & \textbf{0.83}\\
     \hline
    
                    & 8k  &  18.92  &  58.6 & 14.79 & 30.37 & \textbf{3.75} & 4.7 & 4.62\\
    Resampling      & 11k  &  13.73  &  61.41 & 22.17 & 16.51 & 3.63 & \textbf{1.51} & 1.81\\
                    & 22k  &  13.69  &  55.73 & 10.74 & 8.47 & 4.16 & 5.22  & \textbf{2.91}\\
                    & 44k  &  13.68  &  53.3 & 13.51 & 7.4 & \textbf{4.16} & 39.55  & 25.49\\

     \hline
    
    Low Pass Filtering  & 7k  & 13.63 & 50.1 & 10.74 & 10.02 & 3.98 & 5.29 & \textbf{2.83}\\
    
     \hline
     
    \end{tabular}
    \label{table:Laundering_Res} 
\vspace{-1em}
\end{table*}

The performance of all systems under reverberation laundering attack is shown in table \ref{table:Laundering_Res} (row Reverberation). We can observe the decline in systems' performances as the value of reverberation time (RT60) increases. OC-Softmax outperforms other systems under reverberation attack, with an EER of 9.01\%, 15.06\%, and 22.04\% for parameter RT60 $\in$ (0.3s, 0.6s, 0.9s), respectively. In general, LFCC based systems perform better than other systems.

The results of the selected systems in the presence of additive noise attack (Babble, Volvo, White, Cafe, and Street) are shown in table \ref{table:Laundering_Res} (rows 5-9, parameter avg). It can be observed that RawNet2 outperforms RawGat-ST and AASIST under all types of additive noise attack, except Volvo, with an avg EER (average over 0dB, 10dB, and 20dB) of 13.39\% for Babble noise, 6.42\% for Volvo noise, 9.48\% for White Noise, 12.75\% for Cafe noise, and 16.55\% for street noise. In the presence of Volvo noise attack, AASIST outperforms other systems with an avg EER of 5.92\%. 

The performance of the selected systems in the presence of recompression laundering attack is shown in table \ref{table:Laundering_Res} (row Recompression). It can be observed that the selected systems perform worse when compression bit-rate is 16 kbit/s, however for other compression rates, end-to-end learning systems and LFCC-LCNN has the same performance as on ASVSpoof 2019 LA eval. Moreover, we can observe that the EER of RawNet2 is lower for compression rates of 128, 192, 256, and 320 kbit/s. After examining the scores for clean audio files and recompression audio files, we found that some samples have scores equal to 0 (in the order of $10^{-5}$). These borderline cases can cause small differences in the computation of EER either positive or negative. This suggests the need for a detailed analysis, which will be considered in our future work.

In the presence of resampling laundering attack, RawNet2 shows stable performance at all sampling rates. Both RawGat-ST and AASIST are vulnerable to resampling laundering attacks, specifically when the sampling rate is 44KHz, exhibiting an EER of 39.55\% and 25.40\% respectively. One might logically anticipate that the performance of audio spoof detection would approximate that of the unaltered data at sampling rates near the clean data's rate, specifically 16kHz, and that performance would deteriorate as the sampling rate diverges. In fact, this pattern is only observed in methods based on end-to-end learning. However, other approaches do not conform to this expected trend.

\section{Conclusion and Future Work}
This paper evaluated seven audio spoof detection systems in the presence of reverberation, additive noise, recompression, resampling and low-pass filtering laundering attacks. We demonstrate that LFCC features based systems perform better than other systems under reverberation attacks. For other laundering attacks, end-to-end learning systems outperform representation learning and conventional machine learning systems. Additionally, audio spoof detection systems do not follow any trend for recompression and resampling as the recompression bit rates or sampling rates are changed respectively. This suggests the need for a detailed analysis, which we will consider in our future work. Moreover, post processing operations occurring in social and web platforms will also be considered in the future work which, as mentioned earlier, might leave peculiar traces in the audio signals. Furthermore, it would be interesting to train audio spoof detection systems on ASVSpoof Laundered Database and evaluate them on an in-the-wild database to see if the performance improves on an in-the-wild dataset.

\bibliographystyle{ACM-Reference-Format}
\bibliography{references}

\end{document}